\documentclass[usenatbib,onecolumn]{mnras}

\usepackage{graphicx}
\usepackage{amsfonts}
\usepackage{amsmath}
\usepackage{amssymb}
\usepackage{graphicx}
\usepackage{color}
\usepackage[T1]{fontenc}
\usepackage{ae,aecompl}
\usepackage{bm}
\usepackage{pdflscape}
\usepackage[a4paper]{geometry}

\numberwithin{equation}{section}

\title[Normal modes in two-fluid quantum plasmas]{Normal Modes in Magnetized Two-Fluid Spin Quantum Plasmas}

\author[D. G\'omez and A. Kandus]{Daniel O. G\'omez,$^1$\thanks{dgomez@df.uba.ar}
and Alejandra Kandus,$^2$\thanks{kandus@uesc.br}\\
$^1$ Departamento de F\'\i sica, FCEyN - UBA and IAFE (UBA - CONICET), (1428) Ciudad Universitaria, CABA, Argentina.\\
$^2$LATO - DCET - UESC - Rodovia Jorge Amado km 16 s/n,
 (45.662-900) Ilh\'{e}us - BA, Brazil}

\begin{document}

\maketitle

\begin{abstract}
We extend the classical two-fluid magnetohydrodynamic (MHD) formalism to include quantum effects such as electron Fermi pressure, Bohm pressure and spin couplings. 
At scales smaller than the electron skin-depth, the Hall effect and electron inertia must be taken into account, and can overlap with the quantum effects. 
We write down the full set of two-fluid quantum MHD (QMHD) and analyze the relative importance ofthese effects in the high density environments of neutron star atmospheres and white dwarf interiors, finding that for 
a broad range of parameters all these effects are operative. Of all spin interactions we analyze only 
the spin-magnetic coupling, as it is linear in $\hbar$ and consequently it is the strongest spin effect. We re-obtain the classical 
two-fluid MHD dispersion relations corresponding to the magnetosonic and Alfv\'en modes, 
modified by quantum effects. In the zero-spin case, for propagation parallel to the magnetic field, we find that the frequency of the fast mode
is due to quantum effects modified by electron inertia, while the frequency of the Alfv\'en-slow sector
has no quantum corrections. For perpendicular propagation, the fast-mode frequency is the same as for the parallel propagation plus
a correction due only to classical two-fluid effects. When spin is considered, a whistler mode appears, which is due to two-fluid effects
plus spin-magnetic interaction. There are no modifications due to spin for parallel propagation of magnetosonic and Alfv\'en waves, 
while for perpendicular propagation a dispersive term due to spin arises in the two-fluid expression for the fast magnetosonic mode.
\end{abstract} 


\section{Introduction}\label{int}

Magnetohydrodynamics (MHD) can be described as a 'coarse grained' formalism, suitable to study magnetized plasmas at scales larger than
the ion inertial length, $\lambda_i = c/\omega_{pi}$, with $c$ the speed of light and $\omega_{pi}$ the ion plasma frequency. At those
scales, for example, hydrodynamic and magnetohydrodynamic (MHD) turbulence display the same power-law spectrum for the energy cascade,
i.e. a $k^{-5/3}$ Kolmogorov spectrum \citep{Math-Gold-82,Leam-Math-00,Smi-Mull-01}. However, while at shorter scales the hydrodynamic
turbulence still displays a Kolmogorov spectrum all the way down to the dissipation scale, MHD turbulence does not. At scales smaller than $\lambda_i$, 
a regime known as Hall-MHD is established, in which the energy power spectrum becomes somewhat steeper \citep{Gold-Rob-94,Gho-Sir-96,Smi-Ham-06}.  
At scales much smaller than $\lambda_i$, an approximate description known as electron MHD (EMHD) has been proposed, which assumes the ions to 
be static (because of their much larger mass) and consistently the electric currents are fully carried by the electrons. \citet{bisk-99}
studied numerically the EMHD turbulence and confirmed the steeper behavior of the energy spectrum at small scales. Recently \citet{gomez-14a,gomez-14b}
wrote down a complete two-fluid MHD model which includes the Hall and electron inertia effects. Within this description, it is possible to study 
classical plasma effects all the way from scales as large as the size of the system down to the electron inertial scale $\lambda_e=c/\omega_{pe}$, 
and this includes MHD, Hall-MHD and EMHD as asymptotic limits in the appropriate range of scales. At scales below 
$\lambda_{i}$, as a result of the Hall current term, ions are no longer frozen-in to the magnetic field lines, while electrons still remain frozen. 
Therefore, at these scales, the motion of electrons decouple from the one of ions, even though the dynamics can be properly described by Hall-MHD. 
At scales below $\lambda_e$, electrons decouple from magnetic field lines as well, and a proper description of the dynamics at these scales requires 
the full two-fluid MHD description. 

All the plasma effects discussed thus far are classical, and quantum effects will also become relevant at sufficiently small scales. More specifically, 
quantum effects should have to be taken into account whenever the thermal de Broglie wavelength of the plasma particles  $\lambda_B = \hbar/\sqrt{k_B T m}$  
($k_B$: Boltzmann's constant, $\hbar = h/2\pi$, $T$: temperature and $m$: mass) becomes of the order of the interparticle separation, i.e. $\lambda_B > n_0^{-1/3}$ with $n_0$ the average particle density of the plasma. Interparticle separations 
smaller than $\lambda_B$ may be found in extremely dense astrophysical plasmas, such as e.g. white dwarfs, magnetars or 
neutron stars. For example in a magnetar atmosphere we have $T \simeq 10^9$ K and therefore $\lambda_B \simeq 9\times 10^{-11}$ cm which is of the order 
of the interparticle separation. For a white dwarf, $T \simeq 10^4$, and $\lambda_B \simeq 3\times 10^{-8}$ cm which is again of the order of 
$n_0^{-1/3}$. Therefore, it is expected that quantum effects might play a non-negligible role in these extreme astrophysical environments 
\footnote{Other plasma systems where quantum effects might become important are those extremely small, so that the classical transport 
models become invalid. Examples of such systems are nanoscale electronic devices \citep{Cui2001}, thin metal films \citep{SuAl10} and high energy lasers \citep{RidgAl17}.}. 

The formalism to study MHD with quantum effects is known as quantum magnetohydrodynamics (QMHD)
\citep{haas-05,haas-11,mark-brod-07,brod-mark-07b}. In it, the equations of classical MHD are extended to include terms that take into account 
the quantum nature of the charge carriers. The paradigmatic model of a quantum fluid is that of a Fermi gas, with pressure $p_F = (2/5)n_0 E_F = 
\left(3\pi^2\right)^{2/3} \left(\hbar^2/5m\right)n_0^{5/3}$, where $E_F$ is the Fermi energy. 
Besides the Fermi pressur, more complete descriptions include a 'quantum force' whose origin is the Bohm potential due to the overlap of wavefunctions as well as spin effects. These spin effects are mainly due to three sources: a spin-spin coupling due to spin gradients, a spin-density 
coupling due to spin and density gradients and, in the presence of an external magnetic field $\mathbf{B}$, a $spin-\mathbf{B}$ interaction due 
to the coupling of the spins to gradients of the magnetic field. In the presence of inhomogeneous magnetic fields, the most intense 
effect is the $spin-\mathbf{B}$ one, because it is of order $\hbar$ while the others are of order $\hbar^2$.

From what was said in the previous paragraphs, it seems apparent that there might be cases in which  the Hall effect and electron inertia can be as important quantum effects. Therefore, in this manuscript we extend the two-fluid formalism developed by \citet{gomez-14b} by including 
quantum effects such as Fermi pressure, Bohm pressure and spin interactions. Our final aim, is to find a theory that describes as accurately as possible small scale effects in dense magnetized plasmas.

One possible approach to assess how the different effects mentioned in the previous paragraphs affect the dynamical  properties of multispecies 
plasmas is to obtain the dispersion relations for the propagation of linear perturbations of the different quantities that enter in 
the problem (e.g., density, magnetic field, spin). 
For two-fluid plasmas, this study was done by \citet{gomez-14b}. For QMHD we may mention the studies on the propagation 
of linear sonic waves \citep{brod-mark-07a,mark-brod-07,shukla-07,asenjo-12,andreev-15} and of low frequency waves 
\citep{shukla-06a,saleem-08,haas-03}. Non-linear phenomena such as shock waves \citep{misra-08,masood-10} and nonlinear waves \citep{ali-07,shukla-06b}
were also analyzed. Moreover, the effect of radiative processes on quantum plasmas was also addressed \citep{CrReGr-14}. 
This list of references is, of course, not exhaustive. 

We consider an electrically neutral plasma composed by two fermionic fluids of equal modulus and opposite sign charges at temperatures
of the order or below the Fermi temperature. To visualize more clearly the role of the different effects, we neglect kinematic viscosity as 
well as electrical resistivity. In order to analyze the relative importance of each term we rewrite the equations in non-dimensional form 
by defining several dimensionless parameters. This procedure has the advantage of making the analysis independent of the unit system,
avoids spurious over- or under-estimations of the different effects and also allows to directly rescale between completely different systems 
as e.g. the astrophysics and laboratory plasmas (see \citet{CrReGr-14} for a discussion of this procedure).

We write down the system of dimensionless two-fluid  QMHD equations and linearize them around an equilibrium configuration. We find expressions for the dispersion relations of the Alfv\'en and magnetosonic waves that generalize 
the results found previously in the literature on classical MHD \citep{land-ecm} and QMHD \citep{brod-mark-07a} and two-fluid MHD \citep{gomez-14b}. Moreover, due to the presence of a spin-magnetic coupling, we obtain a new dispersion relation that corresponds to a whistler 
mode \citep{stenzel-99}. This mode arises in the two-fluid approach considered, and because of the spin effect it becomes dispersive also at wavelengths 
well larger than the interparticle separation.

The manuscript is organized as follows:
In Section \ref{qmhd} we obtain the two-fluid QMHD equations and analyze the applicability of each effect in the parameter space of astrophysical
compact objects given by $\left(n_0,B_0\right)$, with $B_0$ a mean magnetic field.
In Section \ref{nm} we obtain the generalized dispersion relations for the cases without spin (Subsection \ref{nm-no-spin}) and with spin 
(Subsection \ref{nm-spin}). In Section \ref{conc} we draw our main conclusions. In the Appendix we detail the procedure to turn the 
equations non-dimensional.

We work in c.g.s. units, where  $\hbar = 1.0546\times 10^{-27}\mathrm{cm}^2\mathrm{sec}^{-1}\mathrm{g}$, $c=3\times 10^{10}\mathrm{cm}~\mathrm{sec}^{-1}$,
$m_p = 1.67\times 10^{-24}\mathrm{g}$, $m_e = 9.1\times 10^{-28}\mathrm{g}$, and obtain the electric charge from the fine structure constant $\alpha$, 
i.e. $e^2 \simeq \hbar c/137$. Finally summation over repeated indices is assumed.

\section{Quantum MHD }\label{qmhd}
In this section we derive the QMHD equations for a two-fluid ion-electron plasma in an external magnetic field $\bar B$, starting from the equations 
for each individual species, and analyze the relative importance of each term. 

As stated in Section \ref{int}, we consider an electrically neutral degenerate plasma composed by two species with charges 
$q_s = \pm e$, particle masses $m_s$ and spin $1/2$. The Fermi pressure in three dimensions for a gas of particle mass $m_s$ 
and particle density $n_s$ in the limit $T_s\rightarrow 0$ is
\begin{equation}
p_{s}=\frac{2}{5}n_sE_{Fs}= \left(3\pi^2\right)^{2/3}\frac{\hbar^2}{5m_s}n_s^{5/3}\label{a1-a}
\end{equation}
where $E_{Fs}$ is the Fermi energy of species $s$, given by $E_{Fs} = \left(\hbar^2/2m_s\right)\left(3\pi^2 n_s\right)^{2/3}$.
The equations for each species were considered elsewhere \citep{haas-05,mark-brod-07,brod-mark-07b,mah-as,gomez-14b} and read
\begin{eqnarray}
\partial_t n_s + \bar\nabla\cdot \left(n_s\bar u_s\right) &=& 0 \label{a2-a}\\
\partial_t \bar u_s + \left(\bar u_s\cdot \bar\nabla \right) \bar u_s &=& \frac{q_s}{m_s}\bar E 
+ \frac{q_s}{m_sc}\bar u_s\times \bar B -\frac{1}{m_s n_s}\bar\nabla p_s 
 + \frac{\hbar^2}{2m_s^2}\bar\nabla\left(\frac{\nabla^2 n_s^{1/2}}{n_s^{1/2}}\right)\nonumber\\
&+&\frac{\hbar q_s}{2m_s^2c}S^s_j\bar\nabla\hat {\bar B}^s_j
+ \frac{\hbar^2}{2m_s^2}\bar\nabla\left(\partial_jS^s_i\partial_jS^s_i\right)
\label{a3}\\
\left(\partial_t + \bar u_s\cdot\bar\nabla\right)\bar S^s &=& \frac{q_s}{m_s c}\bar S^s\times \hat {\bar B}^s\label{a4}
\end{eqnarray}
with $c$ the speed of light and where we defined
\begin{equation}
\hat{\bar B}^s = \bar B + \frac{\hbar c}{2q_s n_s} \partial_j\left( n_s \partial_j \bar S^s\right) \label{a7}
\end{equation}
In the previous expressions the average spin vector field for species $s$, $\bar S^s$, satisfies 
$\bar S^s\cdot \bar S^s = 1$. From eqs. (\ref{a3}) and (\ref{a7}) we see that the spin
introduces three forces, $S^i\bar\nabla \bar B^s_i$,  $S^s_i\partial_j\left( n_s \partial_j  S^s_i\right)$
and $\bar\nabla\left( \partial_i S^s_j\partial_i S^s_j\right)$, which arise
 after the passage from particle to fluid description \citep{holland-93}. The first is due to the interaction
of the spins with an external, inhomogeneous magnetic field $\bar B$, the second is caused by an inhomogeneous magnetization 
created by the spins themselves. The other term, $\partial_i S_j\partial_i S_j$ can be interpreted as a spin pressure,
that vanishes if the spin distribution is homogeneous, i.e., spins are completely aligned (see Ref. \citet{mah-as} for an 
analysis of the importance of this term). These equations must be supplemented with Maxwell equations, which for neutral, 
non-relativistic systems read 
\begin{eqnarray}
\bar\nabla\cdot \bar E &=& 0 \label{a5}\\
\bar J = \frac{c}{4\pi}\bar\nabla \times \bar B &=& \sum_s q_s n_s \bar u_s \label{a6}
\end{eqnarray}

To analyze the relative importance of each term we turn to dimensionless variables. The calculations are done in detail in 
Appendix \ref{ndeq}. Here we only quote the different parameters: 
$\mu=m_e/M$, $\beta_0=\left(3\pi^2\right)^{2/3}\left(\lambda_0 n_0^{1/3}\right)^{2}/5$  (quantum plasma $\beta_0$),  $M = m_e + m_p$,
$V_A=B_0/\sqrt{4\pi M n_0}$, $\lambda_0=\hbar/\left(M V_A\right)$ (de Broglie-Alfv\'en length), $\ell = \lambda_0/L_0$, 
$\varepsilon = c/\left(\omega_M L_0\right)$, $\omega_M = \sqrt{4\pi e^2 n_0/M}$ (plasma frequency). $L_0$ is
an arbitrary length scale that we introduced to make lengths non-dimensional. It can be interpreted as the resolution with which we
look at the system. We choose it as a multiple of the particle separation, namely $L_0 = 10^q n_0^{-1/3}$ with $q\geq 0$.
Note that $\ell$ and $\beta_0$ track quantum effects, while $\mu$ tracks electron inertia. 
In the specific case of a proton-electron plasma, the dimensionless equations of motion for each species become
\begin{eqnarray}
\mu \frac{d\bar u_e}{dt} &=& -\frac{1}{\varepsilon}\left(\bar E + \bar u_e\times\bar B\right) 
- \beta_0\frac{\bar\nabla p_e}{n} + \frac{\ell^2}{\mu} \bar\nabla \left(\frac{\nabla^2 n^{1/2}}{2n^{1/2}}\right) 
- \frac{\ell}{2\varepsilon\mu}S_j^e\bar\nabla\hat B^e_j + \frac{\ell^2}{2\mu}\bar\nabla\left(\partial_jS^e_i\partial_jS^e_i \right)
\label{a20}\\
\left(1 -\mu\right) \frac{d\bar u_p}{dt} &=& \frac{1}{\varepsilon}\left(\bar E + \bar u_p\times\bar B\right) 
- \beta_0\frac{\bar\nabla p_p}{n}
+ \frac{\ell^2}{\left(1-\mu\right)} \bar\nabla \left(\frac{\nabla^2 n^{1/2}}{2n^{1/2}}\right)\nonumber\\
&+& \frac{\ell}{2\varepsilon(1-\mu)}S_j^p\bar\nabla\hat B^p_j 
+ \frac{\ell^2}{2(1-\mu)}\bar\nabla\left(\partial_jS^p_i\partial_jS^p_i \right)
\label{a21}\\
\left(\partial_t + \bar u_e\cdot\bar\nabla\right)\bar S^e &=& - \frac{1}{\mu\varepsilon}\bar S^e\times \hat {\bar B}^e\label{a22}\\
\left(\partial_t + \bar u_p\cdot\bar\nabla\right)\bar S^p &=&  \frac{1}{(1-\mu)\varepsilon}\bar S^p\times \hat {\bar B}^p\label{a23}
\end{eqnarray}
and
\begin{eqnarray}
\hat{\bar B}_{e,p} &=& \bar B \pm \frac{\ell \varepsilon}{2n}\partial_i\left( n\partial_i\bar S_{e,p}\right)\label{a24}\\
\mu p_e &=& \left(1-\mu\right) p_p = n^{5/3}\label{a25}\\
\bar J &=&  \frac{n}{\varepsilon}\left(\bar u_p - \bar u_e\right)\label{a26}
\end{eqnarray}

\subsection{Two-fluid QMHD Equations}\label{qmhd-pl}
To describe the system in terms of single fluid variables, we begin by defining the hydrodynamic velocity field $\bar u$ in the usual way, namely
\begin{eqnarray}
 \bar u &=& \left( 1 - \mu\right)\bar u_p + \mu \bar u_e \label{b1-a}\\
 \frac{\varepsilon}{n}\bar J &=& \bar u_p - \bar u_e\label{b2-a}
\end{eqnarray}
from where we obtain
\begin{eqnarray}
\bar u_p &=& \bar u + \frac{\varepsilon\mu}{n}\bar J\label{b3-a}\\
\bar u_e &=& \bar u - \left(1-\mu\right)\frac{\varepsilon}{n}\bar J\label{b4-a}
\end{eqnarray}
The continuity equation is obtained by adding eq. (\ref{a2-a}) for the two species, i.e.,
\begin{equation}
 \partial_t n + \bar\nabla\cdot\left( n\bar u\right) = 0 \label{a8-b}
\end{equation}
Adding eqs. (\ref{a20}) and (\ref{a21}) we obtain the evolution equation for $\bar u$, namely
\begin{eqnarray}
\frac{d\bar u}{dt} &=& \frac{\bar J}{n}\times\left[\bar B + \left(1-\mu\right)\mu \varepsilon^2\bar\nabla\times
\left(\frac{\bar J}{n} \right)\right] 
- \bar\nabla\left[\left(1-\mu\right)\mu \varepsilon^2\frac{J^2}{2n^2}\right]
-\frac{5}{2}\frac{\beta_0}{\mu\left(1-\mu\right)}\bar\nabla n^{2/3} \nonumber\\
&+& \frac{\ell^2}{\mu\left(1-\mu\right)}
\bar\nabla\left(\frac{\nabla^2\sqrt{n}}{2\sqrt{n}} \right) + \mathrm{``spin ~ forces''}\label{b5-b}
\end{eqnarray}
where by ``spin forces'' we mean the sum of all spin dependent terms. Unlike the other non-linear terms, they cannot 
be written in a compact form. Notwithstanding, this fact will not be a drawback as the main spin effects are due to
electrons.
The remaining equation is eq. (\ref{a20}) with the replacement $\bar E = -\partial_t\bar A - \bar\nabla \phi$. It reads
\begin{eqnarray}
 \partial_t\bar A +\bar\nabla\phi &=& \bar u_e\times\bar B + \varepsilon\mu \frac{d\bar u_e}{dt} + \beta_0\varepsilon
 \frac{\bar\nabla n^{5/3}}{n} 
 - \frac{\ell^2\varepsilon}{\mu}\bar\nabla\left(\frac{\nabla^2\sqrt{n}}{2\sqrt{n}} \right)
 + \frac{\ell}{2\varepsilon\mu}S_j^e\bar\nabla\hat B^e_j \nonumber\\
 &&- \frac{\ell^2}{2\mu}\bar\nabla\left(\partial_jS^e_i\partial_jS^e_i \right)
 \label{b6-b}
\end{eqnarray} 
In the r.h.s. of eq. (\ref{b6-b}), the first term is the Hall effect on the electrons, the second term represents electron inertia, 
the third term is the force due to Fermi pressure, the fourth term is the Bohm force, the fifth is the force due to the spin-magnetic field 
coupling and the sixth is the force due to spin-spin couplings. Note that the sixth term is quadratic in $\ell$. Due to the 
definition of $\hat B$ (equation (\ref{a7})), the fifth term consists of two contributions: one due to the coupling of the spin
with the external magnetic field which is linear in $\ell$, and another due to the coupling of the spin with density 
and spin gradients which is quadratic in $\ell$. This means that the main spin contribution comes from the coupling between an external,
inhomogeneous magnetic field and spin, unless spin gradients are strong enough to compensate for the smallness of $\hbar$.
Taking the curl of eq. (\ref{b6-b}) eliminates all `gradient' forces and gives rise to a generalized induction equation:
\begin{equation}
\partial_t\bar B = \bar\nabla\times \left( \bar u_e\times\bar B  \right) + \varepsilon\mu \frac{d}{dt}
\left(\bar\nabla\times\bar u_e\right) + \frac{\ell}{2\varepsilon\mu}\bar\nabla S_j^e\times\bar\nabla\hat B^e_j
\label{b7-b}
\end{equation}
The first two terms are the classic ones of MHD (actually Hall MHD, since the Hall term is included), the second term in the r.h.s. represents a 
battery effect due to electron inertia while the last one is a spin electromotive force.

\subsection{Parameters space}

Due to their much smaller mass, it is likely that the electrons will be responsible for the main dynamical effects. Therefore
in order to assess the relative importance of the different terms in eq. (\ref{b6-b}), we shall plot the coefficient of the 
different terms as functions of the background magnetic field $B_0$ and background density $n_0$. We recall that $L_0 = 10^q n_0^{-1/3}$ 
with $q\geq 0$, represents the resolution with which we observe the plasma. The number of particles in each volume element of linear 
size $L_0$ is then $\mathcal{N}=n_0L_0^3=10^{3q}$. Our parameter space is $\left( n_0, ~ B_0\right)$.

\paragraph{Hall effect:} 
It is described by $\log \varepsilon = 7.4 - q - (1/6)\log n_0$. The region in parameter space such that $\log \varepsilon < 0$
means negligible Hall effect in comparison to the reference terms, while for $\log \varepsilon \geq 0$ i.e.,
when $\log n_0 \leq 6\left(7.4 - q\right)$, it must be taken into account.
\paragraph{Electron inertia:} 
It must be taken into account when $\log \left( \mu\varepsilon^2\right) = 11.5 - 2q - (1/3)\log n_0 \geq 0$, i.e., when
$\log n_0 \leq 3\left( 11.5 - 2q\right)$.
\paragraph{Fermi pressure:}
It is important for $\log \left(\beta_0\varepsilon\right) \geq 0$, which means $\log B_0 \leq - 11.55 - q/2 + (3/4)\log n_0$.
\paragraph{Bohm pressure:}
It is not negligible when $\log\left( \ell^2\varepsilon /\mu\right) \geq 0$, or equivalently when $\log B_0 \leq -9.25 -
(3/2)q + (3/4)\log n_0$.
\paragraph{Spin forces:}
As mentioned above, spin forces are due to two interactions: one with the
external magnetic field, $S_j\bar\nabla B^j$ with weight $\ell /2\mu$, and the other due to spin and density gradients, 
$S_j\bar\nabla\left[\partial_i\left(n\partial_iS^j\right)\right]$ and $\bar\nabla\left(\partial_jS_i\partial_jS_i\right)$
both with weight $\ell^2\varepsilon /2\mu$. The latter is important when $\log\left(\ell^2\varepsilon /2\mu\right) \geq 0$
which is equivalent to $\log B_0 \leq -19.98 - (3/2)q + (3/4)\log n_0$. The former will play a non-negligible r\^ole 
if $\log\left(\ell/\varepsilon\mu\right) > 0$, which after replacing the different expressions gives $\log B_0 <
-17 \log(3.19) + \log\left(n_0\right)$.
\vskip 0.5cm

\begin{figure}
\centering
 \includegraphics[scale=0.9]{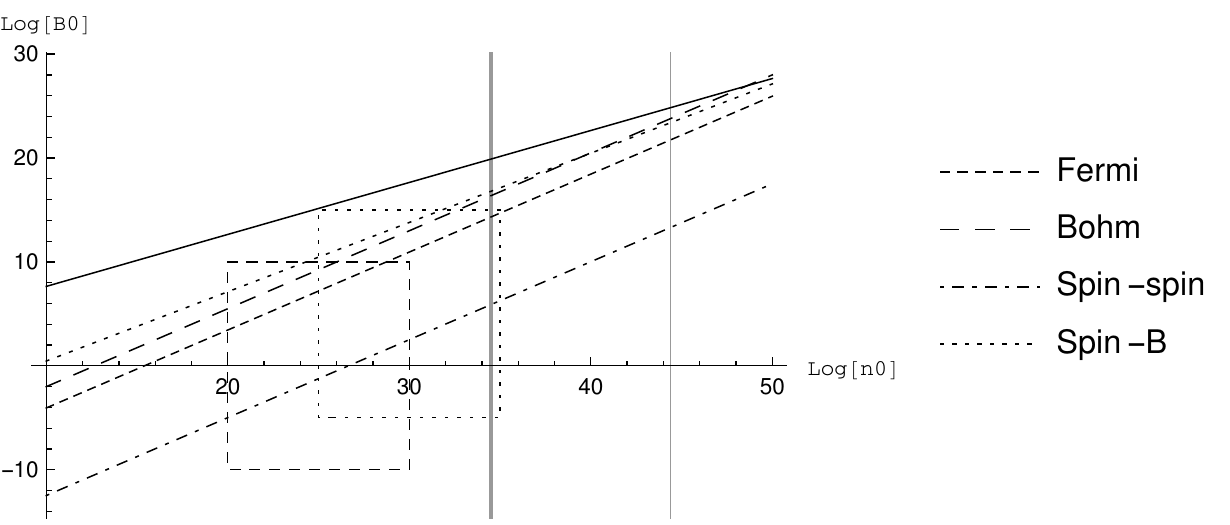} \\
 \caption{\label{param} \textbf{Parameter Space}: Particle density in cm$^{-3}$ and magnetic field in Gauss. Resolution is 
 of the order of the interparticle separation ($q=0$).
 The dashed rectangle shows the region that corresponds to white dwarfs, while the dotted rectangle 
 delimits the parameter's region corresponding to magnetar atmospheres. To the 
 left of the vertical thin line the Hall effect is important, while to the left of the thick vertical line electron inertia must be
 taken into account. Above the continuous oblique line the Alfv\'en velocity becomes larger than $c$, indicating the breakdown
 of the non-relativistic treatment. Therefore, all the effects considered are reasonably well described by the non-relativistic formulation of quantum plasmas.
 Below the dotted oblique line spin-B coupling is important, below the short-dash line Fermi pressure is not
 negligible, below the long-dash line Bohm pressure must be taken into account and below the dashed-dotted line spin-spin coupling becomes
 important.}
\end{figure}

In fig. \ref{param} we have plotted the curves that we have just described, for $q=0$, i.e. the resolution scale is the interparticle
separation. The horizontal axis corresponds to $\log n_0$ and the vertical axis is $\log B_0$. The dashed rectangle shows the parameter space 
region that corresponds to white dwarfs, while the dotted rectangle indicates the parameter's region corresponding to magnetar atmospheres.
The thin vertical line corresponds to the Hall effect: to the left of this line the Hall effect must be taken into account. To the left of the 
thick vertical line, electron inertia is important. We see that for compact astrophysical objects the Hall effect is in general non-negligible, 
while electron inertia will be important for the whole range of densities at scales of the order of the interparticle separation. For poorer resolution, 
i.e. for $q>0$ both vertical lines will be displaced to the left. Below the short-dashed oblique line, Fermi pressure must be taken into account. 
Below the long-dashed line Bohm pressure plays a non-negligible role and below the dashed-dotted line pure spin forces (i.e., due to the term 
$\partial_iS_j\partial_iS_j$) must be considered. The dotted line indicates the spin-magnetic interaction, i.e. $S_j\partial_i B_j$ and 
$S_j\bar\nabla\left[\partial_i\left(n\partial_iS^j\right)\right]$. Below this line, these terms must be taken into account. Above the continuous 
oblique line the Alv\'en velocity becomes larger than the speed of light, indicating that relativistic effects become important. The position of 
the different lines below the continuous one indicates that the non-relativistic treament is adequate for the parameters interval considered in this work.
For astrophysical compact objects we see that the terms of eq. (\ref{b6-b}) that should in principle be considered are the Hall effect, $spin-B$ 
interaction, Fermi pressure and Bohm pressure while second order spin terms are important for weak magnetic fields. 

At this point it is important to calculate at which densities the electron skin-depth is smaller than the interparticle 
separation. Replacing the figures quoted at the end of the Introduction in $\lambda_e = c/\omega_{pe}$, we find that 
$\lambda_e \lesssim n_0^{-1/3}$  for $n_0 \gtrsim 1.09\times 10^{27}$ cm$^{-3}$. From Fig. \ref{param} we see that for most of the density range, 
this relation is satisfied, thus confirming that the two-fluid MHD treatment is indeed correct.

\section{Normal modes in two-fluid QMHD}\label{nm}

To find the normal modes we linearize equations (\ref{a8-b}), (\ref{b5-b}) and (\ref{b7-b}) around a homogeneous equilibrium 
configuration and transform Fourier in space and time. In order to have a better understanding of how Fermi and Bohm 
pressures modify the behavior of the standard normal modes, we  disregard spin effects in a first analysis. 
We shall take them into account in Subsection \ref{nm-spin} and compare the differences they introduce in the behavior of the 
spinless modes. We shall consider only the effect of the coupling of spin to an external magnetic field, in view of the 
fact that it is linear in $\hbar$. Moreover, due to the high non-linearity of the other spin effects, of order $\hbar^2$, 
their main contribution will be on modes of extremely short wavelength, where the fluid assumption might eventually break down.

\subsection{Normal modes without spin forces}\label{nm-no-spin}
Without loss of generality, we consider an equilibrium configuration given by $\langle\bar B\rangle =  \check z$, $\langle\bar u\rangle = 0$ and 
$\langle n\rangle = 1$ and wave vector
$\bar k = k\left(\sin\theta , 0, \cos\theta\right)$. 
Linear perturbations around the equilibrium configuration are
$\bar b = b_{\perp}\left(\cos\theta , 0, -\sin\theta\right) + b_y\left(0 , 1, 0\right)$ and
$\bar u = u_{\perp}\left(\cos\theta , 0, -\sin\theta\right) + u_y\left(0 , 1, 0\right) +
u_{\shortparallel}\left(\sin\theta , 0, \cos\theta\right)$. Note that $\bar k\cdot \bar b = 0$. 
Replacing in eqs. (\ref{a8-b}), (\ref{b5-b}) and (\ref{b7-b}) and defining $v=\omega / k$,
we write the resulting set of equations in matrix form to help to visualize the structure of the modes:

\begin{equation}
\left(\begin{matrix}
v & -1 & 0 & 0 & 0 & 0 \\
-\beta\left(k\right) & v & 0 & 0 & 0 & \sin\theta\\
0 & 0 & v & \cos\theta & 0 & 0 \\
0 & 0 & \cos\theta & v \left[ 1 + \mu\left(1-\mu\right)\varepsilon^2 k^2 \right] & -i\varepsilon k\mu v &
-i\varepsilon \left(1-\mu\right) k\cos\theta\\
0 & 0 & 0 & 0 & v & \cos\theta\\
0 & \sin\theta & i\mu\varepsilon k v & i\left( 1 - \mu\right) \varepsilon k \cos\theta & \cos\theta &
v \left[ 1 + \mu\left(1-\mu\right)\varepsilon^2 k^2 \right]
\end{matrix}
\right)
\left(\begin{matrix}
  n\\
  u_{\shortparallel}\\
  u_y\\
  b_y\\
  u_{\perp}\\
  b_{\perp}
\end{matrix}
\right) = 0 \label{d4}
\end{equation}

where
\begin{equation}
 \beta\left(k\right) = \frac{(5/3)\beta_0 + \ell^2k^2/4}{\mu\left(1-\mu\right)}\label{d4-b}
\end{equation}
Eqs. (\ref{d4}) constitute a linear set for the unknowns $n,~ u_{\shortparallel}, ~u_{\perp},~u_y,~ b_{\perp},~
b_y$. These equations contain electron inertia (through $\mu$) and quantum effects (through $\beta_0$ and $\ell$) and
thus extend the standard derivation for the classical magnetosonic and Alfv\'en modes down to scales smaller than the ion 
skin-depth. Observe that $\left(u_{\parallel},~n\right)$ correspond to the fast mode, while $\left(u_y,b_y,u_{\perp},b_{\perp}\right)$ 
correspond to the Alfv\'en-slow sector. For any arbitrary direction of propagation $\theta$ with respect to the equilibrium magnetic field, the determinant of the square matrix in 
eq. (\ref{d4}) gives a sixth order polynomial, $a_6 v^6 + a_4 v^4 + a_2 v^2 + a_0 = 0$ with
\begin{eqnarray}
a_6 &=& \left[1+\mu\left(1-\mu\right)\varepsilon^2 k^2\right]^2 \label{d13-1}\\
a_4 &=& -\cos^2\left(\theta\right)\left[2+k^2\varepsilon^2\left(1-2\mu+2\mu^2\right)\right]
-\left[1+\mu\left(1-\mu\right)\varepsilon^2 k^2\right]^2\beta\left(k\right)
-\sin^2\left(\theta\right)\left[1+\mu\left(1-\mu\right)k^2\varepsilon^2 \right] 
\label{d13-2}\\
a_2 &=& \cos^2\theta \left\{ 1 + \beta\left(k\right)\left[ 2 + k^2\varepsilon^2\left(1-2\mu
+ 2\mu^2\right) 
\right] \right\}\label{d13-3}\\
a_0 &=& -\beta\left(k\right)\cos^4\theta  \label{d13-4}
\end{eqnarray}

In order to better understand how the various effects modify the behavior of the different modes, we analyze the asymptotic  
configurations of parallel and perpendicular propagation.

\subsubsection{Parallel propagation}\label{ns-par}

\begin{figure}
\centering
 \includegraphics[scale=0.5]{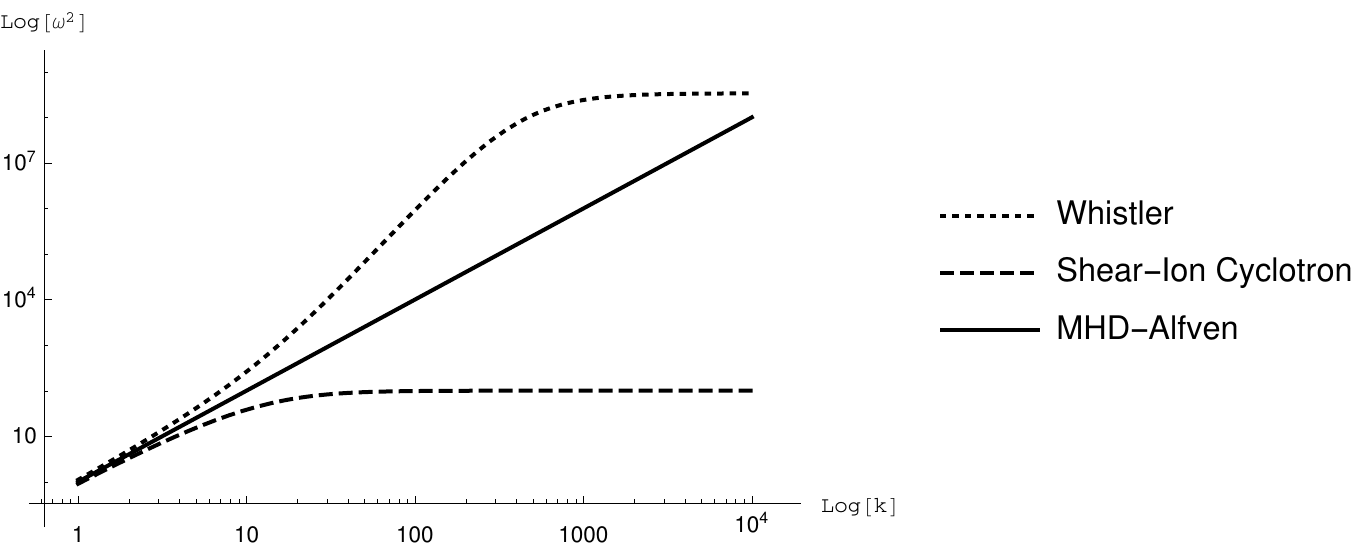} \\
 \caption{\label{nospin-AS} \textbf{No-spin Alfv\'en-Slow sector}: Lower branch corresponds to shear ion-ciclotron waves while upper branch
 to whistler waves. Middle straight line corresponds to the MHD Alfv\'en mode, which is shown for reference. }
\end{figure}

This case is characterized by $\bar k \parallel \bar B_0$, i.e.,
$\theta = 0$. Note that the fast mode decouples from the Alfv\'en-slow sector. This fast mode is described by
\begin{equation}
\left(
\begin{matrix}
 v & -1 \\
 -\beta\left(k\right) & v
\end{matrix}
\right)\left(
\begin{matrix}
n\\
u_{\shortparallel}
\end{matrix}
\right) = 0 \label{d6}
\end{equation}
which corresponds to an acoustic mode with propagation speed $v^2_{\parallel f} = \beta (k)$ given by Eqn. (\ref{d4-b}) above.
Due to Bohm pressure (i.e., $\ell\not=0$), this mode is dispersive for wavenumbers such that
 $\ell^2 k^2/4\gtrsim 5\beta_0/3$, which leads to wavelengths $\lambda = 2\pi/k$ satisfying
 $\lambda \lesssim \left(3\pi^2 \right)^{1/6}10^{-2q} n_0^{-1/3}\sim 1.7 \times 10^{-2q}n_0^{-1/3}$, i.e,  about the 
mean particle separation or smaller. This is consistent with the scale given by the thermal de Broglie wavelength,
indicating that quantum effects are operative at those scales. For the Alfv\'en-slow sector we have

\begin{equation}
\left(\begin{matrix}
v & 1 & 0 & 0 \\
1 & v\left[ 1 + \mu\left(1 - \mu\right)\varepsilon^2 k^2\right] & - i\mu\varepsilon k v & -i\left(1-\mu\right)\varepsilon k \\
0 & 0 & v & 1 \\
i\mu\varepsilon k v & i\left(1-\mu\right)\varepsilon k & 1 & v\left[ 1 + \mu\left(1 - \mu\right)\varepsilon^2 k^2\right]
      \end{matrix}
\right)  \left(\begin{matrix}
                 u_y\\
                 b_y\\
                 u_{\perp}\\
                 b_{\perp}
                \end{matrix}
\right) = 0 \label{d8}
\end{equation}

which gives the following dispersion relation
\begin{equation}
v^4 \left[ 1 + \mu\left(1 - \mu\right)\varepsilon^2 k^2\right]^2 - 
2v^2 \left[ 1 + \frac{\mu^2 +\left(1 - \mu\right)^2}{2}\varepsilon^2 k^2\right] + 1 = 0 \label{d9}
\end{equation}
Note that this equation has no quantum effects. The corresponding solutions were found in \citet{gomez-14a} and we refer
the reader to that reference for the detailed analysis of the corresponding modes. It is clear that the difference with the 
MHD dispersion relation is due to the two-fluid effects considered. The two main roots of eq. (\ref{d9}) are shown in Fig. 
\ref{nospin-AS}. At high wavenumbers, the MHD Alfv\'en frequency gives rise to two modes, a shear ion-cyclotron mode (lower branch) 
and a whistler mode (upper branch). The latter saturates at the electron cyclotron frequency, while the former does so 
at the proton cyclotron frequency \citep{gomez-14a}.

\subsubsection{Perpendicular propagation}\label{ns-perp}

For $\bar k\perp \bar B_0$, i.e., $\theta = \pi/2$, we see from expr. (\ref{d4}) that the subspace $\left( b_y,~ u_{\perp} \right)$
becomes degenerate with $\left( u_y,~ b_{\perp} \right)$. For the remaining variables we have
\begin{equation}
\left(
\begin{matrix}
 v & -1 & 0 & 0 \\
 -\beta\left(k\right) & v & 0 & 1\\
 0 & 0 & v & 0\\
 0 & 1 & i\mu\varepsilon k v & v\left[ 1 + \mu\left(1 - \mu\right)\varepsilon^2 k^2\right]
\end{matrix}
\right)\left(
\begin{matrix}
 n \\
 u_{\shortparallel}\\
 u_y\\
 b_{\perp}
\end{matrix}
\right) = 0 \label{d10}
\end{equation}
which leads to the following dispersion relation
\begin{equation}
 v^2\left[1 + \mu\left(1 - \mu\right)\varepsilon^2 k^2\right] - \left[ 1 +
 \beta\left(k\right)\left( 1 + \mu\left(1-\mu\right)\varepsilon^2 k^2\right)
 \right]=0\label{d11}
\end{equation}
that reduces to the following propagation speed for the fast mode
\begin{equation}
v^2_{\perp f} = \frac{1}{1 + \mu\left(1-\mu\right)\varepsilon^2 k^2} + \frac{5\beta_0/3 + \ell^2 k^2/4}{\mu\left(1-\mu\right)} 
\label{d12b}
\end{equation}
We see that the acoustic fast mode dispersion relation is modified by electron inertia ($\mu\varepsilon^2$) and by quantum effects
($\beta_0,~\ell^2$).
In view of the discussion on the fast mode made in the previous subsection, we know that the effect of the Bohm pressure is important
for scales of the order of the interparticle separation or smaller. To estimate the importance of the first term in (\ref{d12b})
for scales larger than $n_0^{1/3}$ we neglect the correction $\propto \ell^2$ in expr. (\ref{d12b}). Hence the first term will surpass 
the second one for modes such that $k^2 < \left[ 3/5\beta_0 - 1/\mu\left(1-\mu\right)\right]/\varepsilon^2$. However it is easy to check that 
for the most part of the parameter's space of white dwarfs and magnetars the term between square brackets is negative. Only for a small region in the 
left lowest corner of the parameter space, Fig. \ref{param}, i.e. low densities and weak magnetic fields, electron inertia can be operative at any scale.
We therefore conclude that for scales larger than the interparticle separation Fermi pressure is the dominant effect for perpendicular propagation.

\subsection{Inclusion of Spin Effects}\label{nm-spin}

Let us now take into account the effects of spin. We consider only the term proportional to $\bar B$ in Eqs. (\ref{b5-b}) and (\ref{b6-b}) 
because it is linear in $\ell$ (equivalently in $\hbar$) while the terms proportional to $\partial_l\left( n\partial_lS_j\right)$ 
and to $\left(\partial_j S_i\right)^2$ are both of order $\ell^2$, and consequently are expected to play a weaker role. 

\begin{figure}
\centering
 \includegraphics[scale=.7]{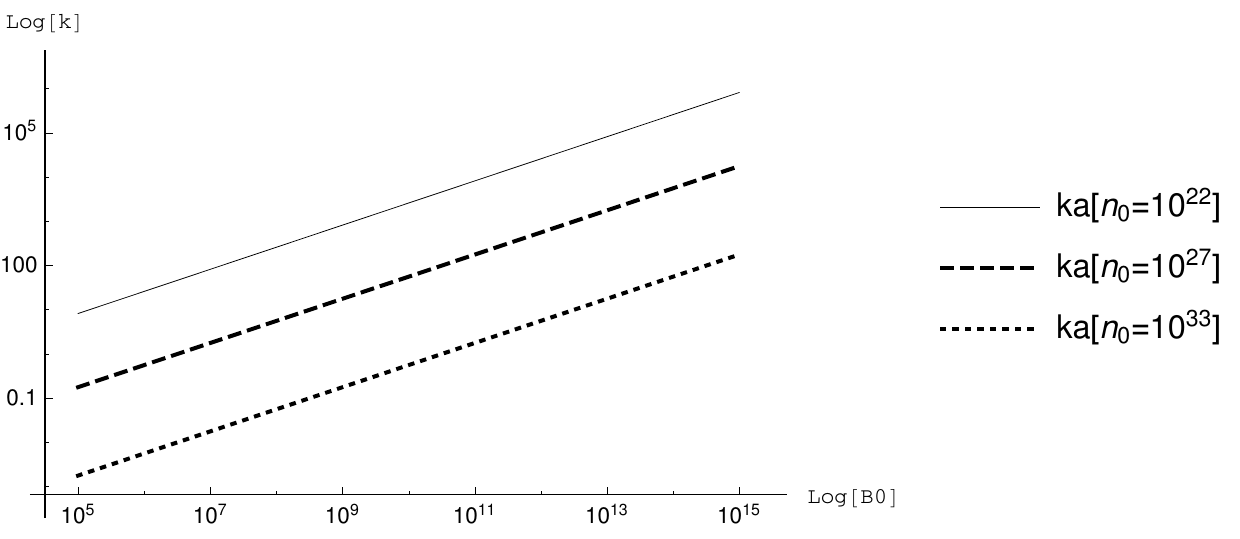} \\
 \caption{\label{whistler-B0} Log-log plot of $k=\sqrt{2/\ell\varepsilon}$ as a function of $B_0$ for three values of the 
 background density. Wavenumbers higher than the ones indicated by the curves experience spin effects.}
\end{figure}

\begin{figure}
\centering
 \includegraphics[scale=.7]{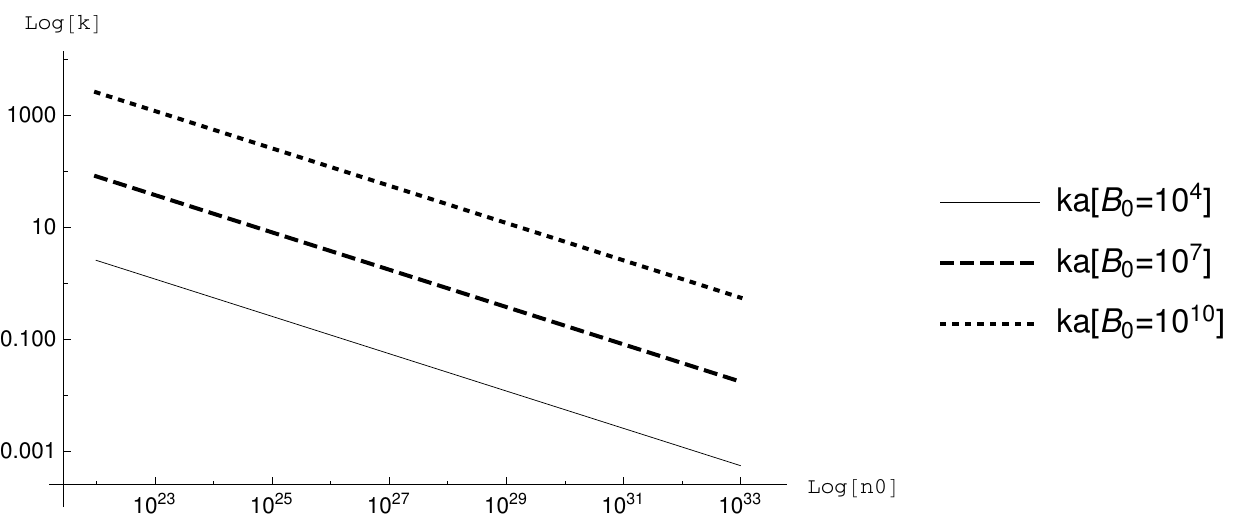} \\
 \caption{\label{whistler-n0} Log-log plot of $k=\sqrt{2/\ell\varepsilon}$ as a function of $n_0$ for three values of the background 
 magnetic field. Wavenumbers higher than the ones indicated by the curves experience spin effects.}
\end{figure}

From the equations of motion for each species (i.e. eqs. (\ref{a20}) and (\ref{a21})), we see that the main contribution to the
spin forces is the one of the electrons, because they are proportional to $\mu^{-1} \gg \left(1 - \mu\right)^{-1}$.
This fact justifies neglecting the ion spin in eq. (\ref{b5-b}). Moreover, as spin forces are gradients,
they disappear from the induction equation, i.e., the curl of eq. (\ref{b6-b}). We must therefore solve the system
\begin{eqnarray}
\frac{d\bar u}{dt} &=& \frac{\bar J}{n}\times\left[\bar B + \left(1-\mu\right)\mu \varepsilon^2\bar\nabla\times
\left(\frac{\bar J}{n} \right)\right] 
- \bar\nabla\left[\left(1-\mu\right)\mu \varepsilon^2\frac{J^2}{2n^2}\right]
-\frac{5}{2}\frac{\beta}{\mu\left(1-\mu\right)}\bar\nabla n^{2/3}\nonumber\\
&+& \frac{\ell^2}{\mu\left(1-\mu\right)}
\bar\nabla\left(\frac{\nabla^2\sqrt{n}}{2\sqrt{n}} \right)
- \frac{\ell}{2\varepsilon\mu}S^e_j\bar\nabla\left[ B_j + \frac{\ell\varepsilon}{2}\partial_l\left(n\partial_lS^e_j\right)\right]  
\label{e1}\\
\left(\partial_t + \bar u_e\cdot\bar\nabla\right) \bar S^e &=& \frac{1}{\mu\varepsilon}\bar S^e \times
\left[\bar B + \frac{\ell\varepsilon}{2}\partial_j\left(n\partial_j \bar S^e\right)\right]\label{e2}
\end{eqnarray}
We now linearize these additional terms around an equilibrium configuration for the spin given by:
\begin{equation}
\bar S^e = \tanh\left(\frac{\mu B_0}{k_B T_e}\right)\hat z + \left(
\begin{matrix}
s_x\\
s_y\\
0
\end{matrix}
\right) \label{e3}
\end{equation}
i.e., we consider deviations in a plane perpendicular to a homogeneous spin configuration along $z$. The function
$\tanh\left(\mu B_0/k_B T_e\right)$ is the Brillouin function that describes a spin distribution in a magnetic field,
in thermodynamic equilibrium at temperature $T_e$. For the present analysis it will
be considered constant and equal to 1. The evolution equations for the velocity perturbations now read
\begin{equation}
\omega \left(
\begin{matrix}
u_{\shortparallel}\sin\theta + u_{\perp}\cos\theta\\
u_y\\
u_{\shortparallel}\cos\theta - u_{\perp}\sin\theta
\end{matrix}
\right) = \dots -\frac{\ell b_{\perp}k\sin\theta}{2\varepsilon\mu}\left(
\begin{matrix}
\sin\theta\\
0\\
\cos\theta
\end{matrix}
\right)\label{e4}
\end{equation}
where with ``$\dots$'' we refer to the terms without spin, already discussed in the previous sections. Eq. (\ref{e2}) leads to
\begin{equation}
-i\omega \left(
\begin{matrix}
s_x\\
s_y\\
0
\end{matrix}
\right) = \frac{1}{\varepsilon\mu}\left(
\begin{matrix}
-b_y + \left( 1 + \ell\varepsilon k^2/2\right)s_y\\
b_{\perp}\cos\theta - \left( 1 + \ell\varepsilon k^2/2\right)s_x\\
0
\end{matrix}
\right)\label{e5}
\end{equation}
The linear problem is now extended from a $6\times 6$ system to an $8\times 8$ one, to accommodate the new spin
variables $s_x,~ s_y$. The full system now is
\begin{equation}
\left(\begin{matrix}
v & -1 & 0 & 0 & 0 & 0 & 0 & 0\\
-\beta & v & 0 & 0 & 0 & \left(1+ \ell k/2\varepsilon\mu \right)\sin\theta & 0 & 0\\
0 & 0 & v & \cos\theta & 0 & 0 & 0 & 0 \\
0 & 0 & \cos\theta & v \left[ 1 + \mu\left(1-\mu\right)\varepsilon^2 k^2 \right] & -i\varepsilon k\mu v &
-i\left(1-\mu\right)\varepsilon k\cos\theta & 0 & 0\\
0 & 0 & 0 & 0 & v & \cos\theta & 0 & 0\\
0 & \sin\theta & i\mu\varepsilon k v & i\left( 1 - \mu\right) \varepsilon k \cos\theta & \cos\theta &
v \left[ 1 + \mu\left(1-\mu\right)\varepsilon^2 k^2 \right] & 0 & 0 \\
0 & 0 & 0 & 1/\varepsilon\mu & 0 & 0 &-iv & -\left(1 + \ell\varepsilon k^2/2\right)/\varepsilon\mu\\
0 & 0 & 0 & 0 & 0 & - 1/\varepsilon\mu & \left(1 + \ell\varepsilon k^2/2\right)/\varepsilon\mu & -iv
\end{matrix}
\right)
\left(\begin{matrix}
  n\\
  u_{\shortparallel}\\
  u_y\\
  b_y\\
  u_{\perp}\\
  b_{\perp}\\
  s_x\\
  s_y
\end{matrix}
\right) = 0 \label{e6}
\end{equation}
The determinant of this $8\times 8$ matrix reduces to
\begin{equation}
\left[v^2 - \left(\frac{1+\ell\varepsilon k^2/2}{\varepsilon\mu}\right)^2\right] \mathbf{\Delta}_6 = 0
\label{e7}
\end{equation}
where $\mathbf{\Delta}_6$ is the determinant of the minor solved in the previous subsection, with the only difference 
with respect to the non-spin case of ``$\left( 1+\ell k/2\varepsilon\mu\right)\sin\theta $'' 
in the $\left( u_{\shortparallel},~b_{\perp}\right)$ coupling. The new factor in Eq. (\ref{e7}) corresponds to a 
whistler wave \citep{stenzel-99}. The  speed of propagation for this whistler is
\begin{equation}
v_{ws} = \pm \left(\frac{1+\ell\varepsilon k^2/2}{\varepsilon\mu}\right)\label{e8}
\end{equation}
and is exclusive of the spin degrees of freedom. This new mode is independent of the one described in the no-spin case, for 
parallel propagation. It is a highly dispersive mode at frequencies higher than the 
electron-cyclotron one. Note that if we set $\ell = 0$ this mode continues to exist, but it is not dispersive and only represents the
transport of a ``spin-label'' due to the spiraling of the electrons around the magnetic lines
and not a real spin effect. This spin effect becomes important for $k>\sqrt{2/\ell\varepsilon}$. In fig.
\ref{whistler-B0} we show this wavenumber as a function of $B_0$ for three different values of $n_0$. We see that 
for weak background magnetic fields and high densities, this mode can be present at long wavelengths. 
In fig. \ref{whistler-n0} we show the dependence of this wavenumber with
$n_0$ for three different values of $B_0$. Once again, we see that the mode can be macroscopic for weak fields and high densities, consistent with the previous plot. According to Fig. (\ref{param}), magnetars would be good candidates for the propagation of this mode.

The dispersion relation for the other three normal modes in QMHD is again of the form $ d_6 v^6 + d_4 v^4 + d_2 v^2 + d_0 = 0$,
with
\begin{eqnarray}
 d_6 &=& \left[1+\mu\left(1-\mu\right)\varepsilon^2 k^2\right]^2\label{e10-1}\\
 d_4 &=& -\cos^2\left(\theta\right)\left[2+k^2\varepsilon^2\left(1-2\mu + 2\mu^2\right)\right]
 - \left[1+\mu\left(1-\mu\right)k^2\varepsilon^2\right]^2\beta\left(k\right)\nonumber\\
 &-&\sin^2\left(\theta\right)\left[1+\mu\left(1-\mu\right)k^2\varepsilon^2\right]\left(1+\frac{\ell k}{2\varepsilon\mu}\right)
 \label{e10-2}\\
 d_2 &=& \cos^2\left(\theta\right) \left\{1 + \beta\left(k\right)\left[2 + k^2\varepsilon^2\left(
 1-2\mu + 2\mu^2\right)  \right] + \frac{\ell k}{2\varepsilon\mu}\sin^2\left(\theta\right)\right\}\label{e10-3}\\
 d_0 &=& - \beta\left(k\right)\cos^4\theta \label{e10-4}
\end{eqnarray}

\subsubsection{Parallel and Perpendicular propagation}
As for the non-spin case, we analyze the parallel and perpendicular propagation
By simple inspection of expressions (\ref{e10-1})-(\ref{e10-4}) we see that there is no spin contribution for parallel propagation 
and consequently the normal modes of this sector coincide with those found in Subsection \ref{ns-par}.

For perpendicular propagation, $d_2 = d_0 = 0$ and again the only surviving mode is the fast mode, modified by the spin-magnetic
coupling, The dispersion relation in this case is
\begin{equation}
v_{\perp s}^2 =  \beta\left(k\right) +\frac{1+\ell k/2\varepsilon\mu}{1+\mu\left(1-\mu\right)\varepsilon^2 k^2}
\label{e14}
\end{equation}
The spin correction will be larger than one if $k> 2\varepsilon\mu/\ell$, or equivalently
$k>\left( cm_e/2\pi e\hbar\right)\left(B_0/n_0\right)\sim 1.8\times 10^{28} \left(B_0/n_0\right)$. For the values of $B_0$
and $n_0$ corresponding to astrophysical compact objects (see Fig. \ref{param}) this corresponds again to wavelengths shorter than the 
interparticle separation. To physically interpret the quantum correction we write
\begin{equation}
\frac{\ell}{2\varepsilon\mu} = \frac{\mu_B B_0 n_0}{B_0^2/4\pi}\label{e14c}
\end{equation}
where $\mu_B = e\hbar/2m_e c$ is the Bohr magneton. Expression (\ref{e14c}) then represents the potential energy density of all the 
electron spins embedded in the external field $B_0$ relative to the magnetic energy density of $B_0$.

\section{Conclusions}\label{conc}

In this paper we extended the two-fluid MHD description of a magnetized plasma developed by \citet{gomez-14b} by including quantum effects such 
as Fermi pressure, Bohm potential and spin interactions. The motivation behind this study is that the mentioned quantum effects are
operative at scales that may overlap with the ones where two-fluid effects must be taken into account, i.e., scales shorter than
the ion-skin depth. At those small scales it is known that the MHD is not an appropriate formalism to describe, for example, the turbulent
energy spectrum, while the inclusion of two-fluid effects such as the Hall effect and electron inertia does account for the small
scales features observed in magnetized plasma turbulence.

We wrote down the complete set of two-fluid QMHD equations and, after linearizing them, we obtained the dispersion relations that generalize 
previous results found in the literature on two-fluid MHD description of linear waves in plasmas \citep{gomez-14b}. We separated our study into spinless and spin 
plasmas and, in each of these cases we analyzed parallel and perpendicular propagation of the linear perturbations.

In the absence of spin effects, we found that for parallel propagation the frequency in the fast magnetosonic sector is determined by
the quantum effects (Fermi pressure corrected by Bohm forces), weighed by the 'non-quantum' two fluid effects. In contrast, the 
frequencies of the Alfv\'en-slow sector depend only on electron inertia and Hall effect, i.e., they are not affected by quantum
effects. The features of this sector were analyzed in detail by \citet{gomez-14a}, who showed that due to the two-fluid effects, 
at high wavenumbers the MHD Alfv\'en mode separates into whistler and ion-cyclotron modes.
For perpendicular propagation on the other side, we only have the fast mode, and we found that its velocity is the one obtained 
for parallel propagation plus a term that depends only on two-fluid effects. For the parameter space of compact objects, this 
'non-quantum' correction, however, is in general not operative, except for very small densities and weak magnetic fields.

We considered spin effects only at the level of the spin-B coupling, as it is linear in $\hbar$. The other spin
interactions, being second order in $\hbar$ are expected to have weaker effects. In this case, irrespective of the propagation
direction, a whistler mode appears, which is exclusive of the spin degrees of freedom. To the extent of our knowledge, until now this mode
was not described in the literature. It arises only when two-fluid effects are considered \citep{gomez-14a,gomez-14b} and, due to the 
electron spin, it is highly dispersive at frequencies higher than the electron-cyclotron one. For high densities, the corresponding wavelengths 
can be well larger than the interparticle separation thus having potentially observable effects. 

For the modes in the fast and Alfv\'en-slow sectors, the presence of spin does not modify parallel propagation of linear modes, 
which retain the same features as they had without spin. For perpendicular propagation, however, the spin-B interaction modifies the fast mode
frequency by introducing a dispersive, spin-dependent correction in the pure two-fluid term which, for the densities and magnetic
field ranges of compact objects, would manifest at scales well smaller than the interparticle separation.

In summary, we extended the two-fluid MHD formalism to include quantum effects and studied the propagation of linear waves. A next step would
be to analyze nonlinear effects, as e.g. shock waves and turbulence, to investigate the modifications that the two-fluid together with
quantum effects introduce in those phenomena.

\section*{Acknowledgements}

D. O. G. acknowledges financial support from grants UBACyT 20020130100629BA to the Department of Physics of FCEyN-UBA and PICT 1007 to IAFE. 
A. K. thanks the Physics Department of Facultad de Ciencias Exactas y Naturales - UBA for kind hospitality during the development of part of this work, 
and also financial support from UESC, BA-Brasil.


\appendix

\section{Making the system of MHD equations non-dimensional}\label{ndeq}

Here we put the equations in dimensionless form. We begin with
\begin{eqnarray}
\partial_t \bar u_s + \left(\bar u_s\cdot \bar\nabla \right) \bar u_s &=& \frac{q_s}{m_s}\bar E 
+ \frac{q_s}{m_sc}\bar u_s\times \bar B -\frac{1}{m_s n_s}\bar\nabla p_s 
+ \frac{\hbar^2}{2m_s^2}\bar\nabla\left(\frac{\nabla^2 n_s^{1/2}}{n_s^{1/2}}\right)
+\frac{\hbar q_s}{2m_s^2c}S^s_j\bar\nabla\hat {\bar B}^s_j \nonumber\\
&+& \frac{\hbar^2}{2m_s^2}\bar\nabla\left(\partial_jS^s_i\partial_jS^s_i\right)
\label{a0}
\end{eqnarray}

We consider the following fiducial quantities (to be properly defined later) to get rid of units: 
$n_0$, $u_0$, $L_0$, $E_0$  and $B_0$. $S^i$ is already dimensionless. Time and spatial
derivatives are then written
\begin{eqnarray}
\nabla &\rightarrow& \frac{1}{L_0}\nabla \label{a1}\\
 \frac{\partial}{\partial t} &\rightarrow& \frac{u_0}{L_0}\frac{\partial}{\partial t} \label{a2}
\end{eqnarray}
We keep the same letters for the dimensionless variables for simplicity. We also write $m_s = M \tilde m_s$. So eq.
(\ref{a0}) becomes
\begin{eqnarray}
\frac{u_0^2}{L_0} \left[\partial_t \bar u_s + \left(\bar u_s\cdot \bar\nabla \right) \bar u_s\right] &=& 
\frac{q_s E_0}{M\tilde m_s}\bar E + \frac{q_s u_0 B_0}{M\tilde m_sc}\bar u_s\times \bar B
-\frac{p_0}{M\tilde m_s n_0 \tilde n_s L_0}\bar\nabla \tilde p_s 
+ \frac{\hbar^2}{2 M^2 \tilde m_s^2L_0^3}\bar\nabla\left(\frac{\nabla^2 \tilde n_s^{1/2}}{\tilde n_s^{1/2}}\right)\nonumber\\
&+&\frac{\hbar q_s}{2M^2 \tilde m_s^2cL_0}S^s_j\bar\nabla\hat {\bar B}^s_j 
+ \frac{\hbar^2}{2 M^2\tilde m_s^2 L_0^3}\bar\nabla\left(\partial_jS^s_i\partial_jS^s_i\right)
\label{a0-a}
\end{eqnarray}
We rewrite it as
\begin{eqnarray}
\tilde m_s\left[\partial_t \bar u_s + \left(\bar u_s\cdot \bar\nabla \right) \bar u_s\right] &=&
\frac{L_0}{u_0^2} \frac{q_s E_0}{M}\bar E +\frac{L_0}{u_0} \frac{q_s B_0 }{Mc}\bar u_s\times \bar B
-\frac{L_0}{u_0^2} \frac{p_0}{M n_0 \tilde n_s L_0}\bar\nabla \tilde p_s 
+ \frac{L_0}{u_0^2} \frac{\hbar^2}{2 M^2 \tilde m_sL_0^3}\bar\nabla\left(\frac{\nabla^2 \tilde n_s^{1/2}}{\tilde n_s^{1/2}}\right)\nonumber\\
&+&\frac{L_0}{u_0^2} \frac{\hbar q_s B_0}{2M^2 \tilde m_scL_0}S^s_j\bar\nabla\hat {\bar B}^s_j 
+ \frac{L_0}{u_0^2} \frac{\hbar^2}{2 M^2\tilde m_s L_0^3}\bar\nabla\left(\partial_jS^s_i\partial_jS^s_i\right)
\label{a0-b}
\end{eqnarray}
For a proton-electron plasma, we define 
\begin{eqnarray}
M &=& m_e + m_p\label{b1}\\
\tilde m_e &=& \mu = \frac{m_e}{M}\label{b2}\\
\end{eqnarray}
and besides
\begin{eqnarray}
E_0 &=& \frac{u_0}{c}B_0\label{b3}\\
u_0 &=& V_A =\frac{B_0}{\sqrt{4\pi M n_0}}\label{b4}\\
p_0 &=& \frac{\left(3\pi^2\right)^{2/3}\hbar^2}{5 M}n_0^{5/3}\label{b5}\\
\omega_M &=& \sqrt{\frac{4\pi e^2 n_0}{M}}\label{b6}\\
\lambda_0 &=& \frac{\hbar}{MV_A}\label{b7}\\
\ell &=& \frac{\lambda_0}{L_0}\label{b8}
\end{eqnarray}
Replacing in (\ref{a0-b}) we have for the electron fluid
\begin{eqnarray}
\mu\left[\partial_t \bar u_s + \left(\bar u_s\cdot \bar\nabla \right) \bar u_s\right] &=&
-\frac{L_0 e B_0}{u_0 c M} \left[\bar E +\bar u_s\times \bar B\right]
-\frac{1}{u_0^2} \frac{p_0}{M n_0}\frac{\bar\nabla \tilde p_s}{\tilde n_s}
+ \frac{1}{u_0^2} \frac{\hbar^2}{2 M^2 \mu L_0^2}\bar\nabla\left(\frac{\nabla^2 \tilde n_s^{1/2}}{\tilde n_s^{1/2}}\right)\nonumber\\
&&-\frac{1}{u_0^2} \frac{\hbar e B_0}{2M^2 \mu c}S^s_j\bar\nabla\hat {\bar B}^s_j 
+ \frac{1}{u_0^2} \frac{\hbar^2}{2 M^2\mu L_0^2}\bar\nabla\left(\partial_jS^s_i\partial_jS^s_i\right)
\label{a0-c}
\end{eqnarray}
The coefficients of the different terms can be cast as
\begin{eqnarray}
\frac{L_0 e B_0}{u_0 c M} &=& \frac{L_0 e  \sqrt{4\pi M n_0 }}{ c M}
= \frac{L_0}{c}\omega_M \equiv \frac{1}{\varepsilon}\label{c1} \\
\frac{1}{u_0^2} \frac{p_0}{M n_0} &=& \frac{1}{V_A^2} \frac{1}{M n_0} \frac{\left(3\pi^2\right)^{2/3}\hbar^2}{5 M}n_0^{5/3}
= \frac{\left(3\pi^2\right)^{2/3}}{5 }\left(\lambda_0 n_0^{1/3}\right)^2
\equiv \beta_0 \label{c2}\\
\frac{1}{u_0^2} \frac{\hbar^2}{2 M^2 \mu L_0^2} &=& \frac{\hbar^2}{V_A^2 M^2} \frac{1}{2  \mu L_0^2} =
\frac{\lambda_0^2}{2\mu L_0^2} = \frac{\ell^2}{2\mu}\label{c3}\\
\frac{1}{u_0^2} \frac{\hbar e B_0}{2M^2 \mu c} &=& 
\frac{\hbar}{V_A M} \frac{B_0 e}{2M V_A \mu c} = \frac{\lambda_0}{L_0}\frac{L_0  \sqrt{4\pi e^2M n_0}}{2M  \mu c}
=  \ell \frac{L_0\omega_M}{2\mu c} = \frac{\ell}{2\varepsilon\mu}
\label{c4}
\end{eqnarray}
Substituting all these expressions into equation (\ref{a0-c}) we obtain
\begin{eqnarray}
\mu\left[\partial_t \bar u_s + \left(\bar u_s\cdot \bar\nabla \right) \bar u_s\right] &=&
-\frac{1}{\varepsilon} \left[\bar E +\bar u_s\times \bar B\right]
-\beta\frac{\bar\nabla \tilde n_s^{5/3}}{\tilde n_s}
+ \frac{\ell^2}{2\mu} \bar\nabla\left(\frac{\nabla^2 \tilde n_s^{1/2}}{\tilde n_s^{1/2}}\right)
-\frac{\ell}{2\varepsilon\mu} S^s_j\bar\nabla\hat {\bar B}^s_j \nonumber\\
&&+ \frac{\ell^2}{2\mu}\bar\nabla\left(\partial_jS^s_i\partial_jS^s_i\right)
\label{a0-d}
\end{eqnarray}

\end{document}